\documentclass[aps,prb,floatfix,reprint]{revtex4-1}
\usepackage[utf8]{inputenc}
\usepackage{amsmath,amsbsy,amssymb,amsfonts}
\usepackage{graphicx, color}
\usepackage{esint}
\usepackage{empheq}
\usepackage{hyperref}
\usepackage{appendix} 

\makeatletter

\usepackage{subfigure}
\usepackage{soul}
\usepackage{xcolor}
\usepackage{soul}
\usepackage{amsthm}
\usepackage{color}
\usepackage{bm}

\makeatother

\begin{document}

\title{Nonlinear Defect Theory of Thermalization in Complex Multimoded Systems}

\author{Emily Kabat$^{1}$, Alba Y. Ramos$^{2,3,*}$, Lucas J. Fern\'andez-Alc\'azar$^{2,3}$,  Tsampikos Kottos$^{1}$\\}

\affiliation{
$^{1}$Wave Transport in Complex Systems Lab, Department of Physics, Wesleyan University, Middletown, CT-06459, USA\\
$^{2}$Institute for Modeling and Innovative Technology, IMIT (CONICET - UNNE), Corrientes W3404AAS, Argentina\\
$^{3}$Physics Department, Natural and Exact Science Faculty, Northeastern University of Argentina, Corrientes W3404AAS, Argentina\\
\textit{$*$ email: } albayramos at exa.unne.edu.ar
}

\date{\today}
\begin{abstract}
We show that a single nonlinear defect can thermalize an initial excitation towards a Rayleigh-Jeans (RJ) state in complex multimoded 
systems. The thermalization can be hindered by disorder-induced localization phenomena which drive the system into a metastable RJ state. 
It involves only a (quasi-)isolated set of prethermal modes and can differ dramatically from the thermal RJ. We develop a one-parameter 
scaling theory that predicts the density of prethermal modes and we derive the modal relaxation rate distribution, establishing analogies 
with the Thouless conductance. Our results are relevant to photonics, optomechanics, and cold atoms.
\end{abstract}
\maketitle

\textit{Introduction.-} The complex dynamics of many-body/many-mode systems in response to nonlinear interactions is 
emerging as fundamental to many different fields ranging from physics, chemistry, and quantum information sciences to 
biology and sociology \cite{Winf80,OMA12,WTM10,H07,GP07,NH15}. Approaches that focus on the microscopic behavior 
of such systems fail to provide actionable descriptions. The one theory that has proven powerful is statistical 
mechanics and thermodynamics \cite{Reif65,Callen1985,Reichl98}. However, many branches of science and technology 
have yet to adapt a theory of thermodynamics and statistical mechanics suitable to their field. For instance, the photonics 
community has only recently begun to develop a thermodynamic theory of beam dynamics in nonlinear multimode photonic 
platforms (NMPPs) like multicore/multimode optical fibers \cite{WHC19,RFKS20,MWJC20,ZWHEGC23,WJPKC20}. In these systems, 
even at moderate injected powers, nonlinear interactions dominate the beam evolution by introducing mode-mixing effects 
that redistribute the initial power across individual modes \cite{WWCW22,FGKMP19,PGHSRMC14,N11,NST19}. Such complex many
-mode configurations are most naturally described by a statistical framework.

In the few years since its conception \cite{WHC19}, optical thermodynamics (OT) has proven extremely successful at predicting 
the modal power distribution of NMPPs.  The key tenant of the OT theory, the prediction of a thermal optical state taking 
the form of a Rayleigh-Jeans (RJ) power distribution, has been observed by various experimental groups using different NMPPs 
\cite{BFKGRMP20,PSWBWCW22,WWCW22,MWJKCP23,BGFBMKMP23}. Such developments hold promise for a variety of technological applications 
including the design of efficient cooling schemes for high-power sources (lasers) \cite{WCW17,WWCW22,KFRSK24,china,ILP12} or new fiber structures 
for medical imaging (endoscopy) \cite{BWF19,PTC15,MLLF12} and fiber optic communications \cite{RFN13,HK13}. Another field that could benefit from developments in OT and photonics is the area of cold atoms \cite{ACBG22,SDC22}. 

All current efforts are focusing on the analysis of thermalization in systems with spatially distributed nonlinearities, 
overlooking the fundamental scenario of one nonlinear impurity. To start with, is one nonlinear defect capable of causing 
thermalization? If so, what are the timescales of such a process? How might the complexity/disorder of 
the underlying linear structure impact the power redistribution of an initial excitation?

Here, we demonstrate that even one nonlinear defect can lead to thermalization of an initial beam excitation towards RJ. We 
derive the statistics of the modal relaxation rates and establish an analogy with the Thouless conductance describing transport 
in mesoscopic structures. In chaotic/ergodic systems \cite{S99,H01,E96}, the distribution is Porter-Thomas, indicating a cohesive 
relaxation behavior of all modes and the suppression of large relaxation rates. As Anderson localization effects due to disorder 
become dominant \cite{A58,SSC13,MB21}, the distribution shifts towards a log-normal with a group of fast relaxing modes separated 
from the rest. These modes play a prominent role by enforcing a two-stage thermalization process as they first converge to a 
metastable RJ state that differs, sometimes dramatically, from the thermal state. In the case of spatial symmetries, the mode 
separation can be more extreme: for example, in 1D periodic lattices we observe a bimodal distribution of fast and slow (even 
vanishing) relaxation rates. Using ideas from Renormalization Group theory\cite{AALR79} (RGT) we describe the density of prethermal 
modes $\mathcal{N}$ by a first-order nonlinear differential equation that takes the simple universal form 
\begin{equation}
    \frac{\partial \mathcal{N}}{\partial \ln N}=\beta(\mathcal{N}),
    \label{betafunction}
\end{equation}
where $N$ is the total number of modes of the system. The above equation implies that the logarithmic derivative is 
a function of $\mathcal{N}$ alone.


\begin{figure}[tb]
    \centering
    \includegraphics[width=0.5\textwidth]{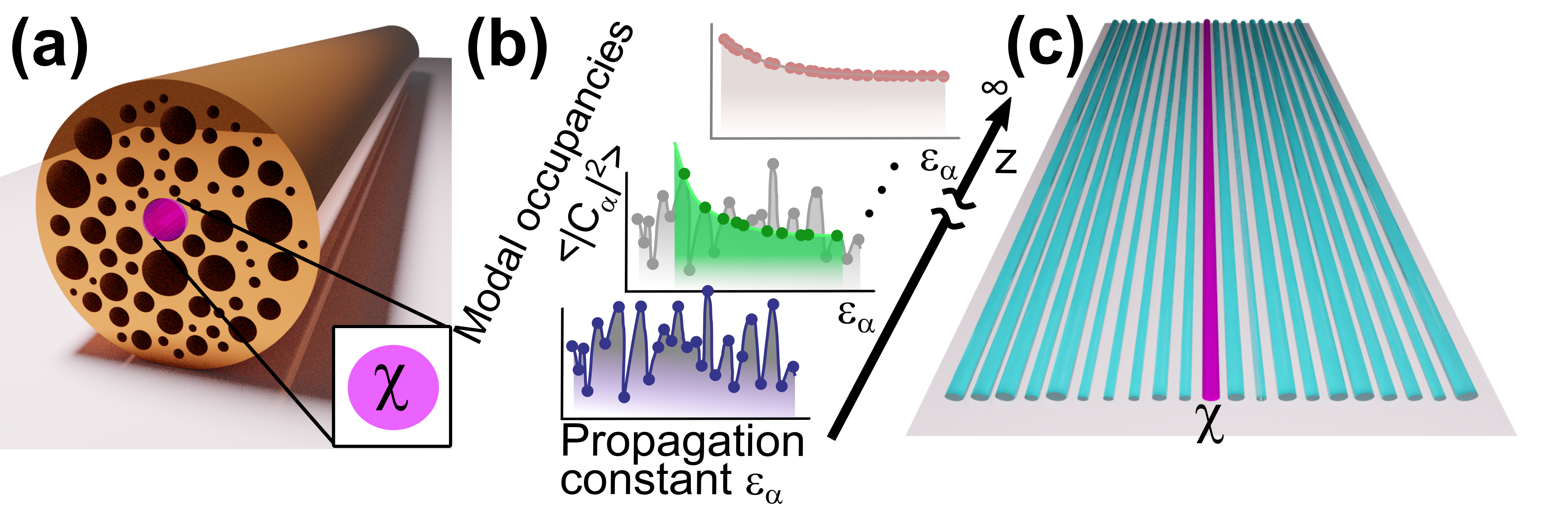}
    \caption{Schematics of $(a)$ a multicore fiber with 
    transverse random long-range coupling ($J_{nl}$) and $(c)$ a 1D array of waveguides with disordered propagation constants. 
    The magenta core/waveguide indicates the position $m$ of the nonlinear defect with strength $\chi$. (b) An 
    initial preparation of mode occupancies $\langle|C_\alpha(z=0)|^2\rangle$ (blue) evolves toward a prethermal state (green) 
    characterized by a prethermal RJ distribution composed of only a fraction of the modes. At very long propagation distances 
    ($z\rightarrow \infty)$, the system thermalizes to a RJ distribution (red) incorporating all modes.}
    \label{fig1}
\end{figure}

\textit{Theoretical framework.-} We consider beam propagation in one-dimensional (1D) and quasi-1D photonic networks \cite{MB21,AM16,LAPSMCS08}, 
Figs. \ref{fig1}(a,c). The beam dynamics is described by a coupled mode theory
\begin{equation}\label{NLS_eq}
i\frac{d\psi_l}{dz}=\sum_n J_{ln}\psi_n+\chi\delta_{lm}|\psi_l|^2\psi_l, \quad l=1,\ldots, N,
\end{equation}
where $\psi_l$ describes the complex field amplitude at the $l-$th core/waveguide, and the time-like variable $z$ is the paraxial 
propagation distance. The last term encodes a Kerr nonlinearity, with $\chi$ being the strength of the nonlinear defect at ``site" $m$. 

The connectivity matrix $J$ determines the geometry of the system, where nondiagonal elements $J_{ln}$ describe the coupling between sites $l$ 
and $n$, and the $J_{l,l}$'s correspond to the propagation constants associated with each core/waveguide. The set-up in Fig. \ref{fig1}(a) can be 
modeled by a Banded Random Matrix (BRM) \cite{Iz90,CMI90,FM91} whose elements $J_{ln}$ are given by a normal Gaussian distribution with mean zero 
$\left\langle J_{nl}\right\rangle=0$ and variance $\left\langle J_{nl}^2\right\rangle=\frac{N+1}{b\left(2N-b+1\right)}$ for $|n-l|< b$ and $J_{nl}=0$ 
for $|n-l|\geq b$. This normalization guarantees that the Hamiltonian (internal energy) associated with Eq. (\ref{NLS_eq}) is extensive. The parameter 
$b$ defines the long-range coupling and can induce Anderson localization of the linear supermodes with localization length $\ell_\infty \sim b^2$. The 
limit $b\rightarrow N$ corresponds to the GOE matrices used in Ref. \cite{RFKS20} for the analysis of thermalization of wave chaotic systems via 
spatially extended nonlinearities. We also consider the 1D case of Fig. \ref{fig1}(c), modeled by a 
nearest neighbor (NN) connectivity matrix with $J_{l,l\pm1}=1$ and random $J_{ll}\in [-\frac W2, \frac W2]$ given by a box distribution. The 
localization length \cite{Economou79,KMK93} of the supermodes is given by $\ell_\infty\approx 24(4-\varepsilon^2)/W^2$.

We represent Eq. (\ref{NLS_eq}) using the supermode basis $f_\alpha(l)$, where $l$ refers to the site index and $\alpha$ the mode index. Then, 
$\psi_l(z)=\sum_\alpha e^{-i \varepsilon_\alpha z}C_\alpha (z)f_\alpha(l)$, and Eq. (\ref{NLS_eq}) reads 
\begin{equation}\label{mode_eq}
    i\frac{dC_\alpha}{dz}=\chi \sum_{\beta \gamma \delta}Q_{\alpha \beta \gamma \delta}C^*_\beta
    C_\gamma C_\delta e^{i\left(\varepsilon_\alpha+\varepsilon_\beta-\varepsilon_\gamma-\varepsilon_\delta\right)z},
\end{equation}
where $\varepsilon_\alpha$ is the $\alpha-$eigenvalue. The RHS represents the mode-mode interactions, where  
\begin{equation}\label{gamma_eq}
     Q_{\alpha \beta \gamma \delta}
     =f^*_\alpha(m)f^*_\beta(m)f_\gamma(m)f_\delta (m)
\end{equation}
determines the strength of the four-wave mode mixing introduced by the nonlinear defect. 

In general, the rate of power exchange between various modes is determined by two processes: the degree of phase matching ($\varepsilon_\alpha+
\varepsilon_\beta-\varepsilon_\gamma-\varepsilon_\delta\approx 0$) and the strength of the mode-mode interaction. When $Q_{\alpha\beta
\gamma\delta}$ is structureless, as in chaotic wave systems where the amplitudes $f_\alpha(m)$ are comparable 
across all $\alpha$-eigenmodes, the phase matching mechanism is the only means by which our systems can achieve thermalization. 
When the degree of mode overlap $f_\alpha(m)$ at the nonlinear defect $m$ is $\alpha$-dependent, $Q_{\alpha\beta\gamma\delta}$ 
acquires a structure that becomes relevant for the analysis of the modal power distribution $|C_\alpha(t)|^2$. Specifically, 
modes that have a high amplitude at the nonlinear site have the potential to be involved in many significant four-wave mode 
interactions. Conversely, modes with small amplitudes are excluded from all four-wave mode interactions. Thus we see 
the emergence of two groups of modes: interacting ones and non-interacting ones. These two groups act as (quasi-)isolated systems 
with constant internal energy and power.

{\it Prethermalized modes and metastable RJ --} While standard analysis of the power redistribution in NMPPs involves the numerical 
integration of many coupled differential equations (see Eqs. (\ref{NLS_eq},\ref{mode_eq})), the recent development of optical 
thermodynamics (OT) offers an elegant alternative \cite{WHC19,RFKS20,MWJC20}. This framework posits a thermal equilibrium without 
addressing questions about the thermalization process. It identifies intrinsic variables $T,\mu$ that are optical thermodynamic 
forces analogous to chemical potential and temperature in traditional statistical mechanics. Both are uniquely determined by the 
two constants of motion of Eq. (\ref{mode_eq}): total power ${\cal A}=\sum_\alpha|C_\alpha|^2$ and total internal energy ${\cal H}
\approx{\cal H}_{\rm Linear}=\sum_\alpha \varepsilon_\alpha |C_\alpha|^2$ (assuming weak nonlinearities) \cite{RCKG00}. The thermal RJ distribution 
of modal power takes the form $\langle |C_\alpha|^2\rangle= n_\alpha=\frac{T}{\varepsilon_\alpha-\mu}$ where $\langle\cdot\rangle$ 
indicates thermal averaging. 

\begin{figure}[h!]
    \centering
    \includegraphics[width=0.5\textwidth]{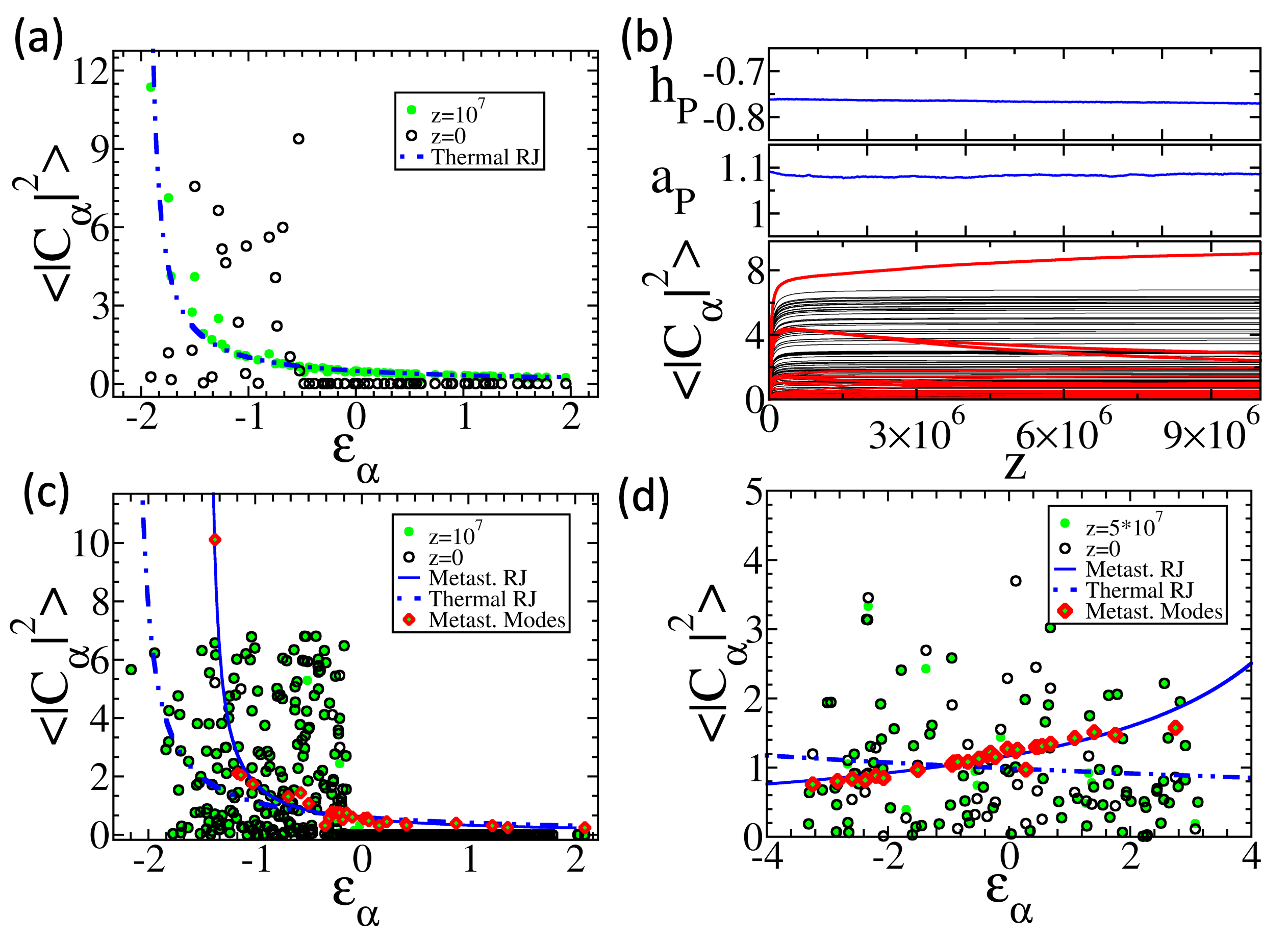}
    \caption{(a) Initial (black circles) and final (green circles) modal occupations for a BRM with $b=N=64$. Due to supermode ergodicity, 
    the entire system reaches a RJ thermal state (dashed blue line) with $(T,\mu)=(1.0,-2.0)$. (b) Evolution of energy density 
    $h_P\equiv {\cal H}_P/N_P$ (up), optical power density $a_P\equiv {\cal A}_P/N_P$ (middle) and modal occupation 
    $\langle|C_\alpha(z)|^2\rangle$ (bottom). The prethermal modes (red) display a long-lived stability after reaching the prethermal 
    RJ state. (c) BRM setting, displaying a metastable RJ state (solid blue line) with $(T_P, \mu_P)=(0.8,-1.5)$ that differs dramatically 
    from the final RJ state (dashed blue line) with $(T, \mu)=(1.3,-2.2)$. Prethermal modes are highlighted in red diamonds. (d) NN 
    equivalent of (c), with $(T_P, \mu_P)=(-8.8,7.5)$ and ($T, \mu)=(24.9,-25.2)$. In (b-c) $N=512$, $b=3$ and $m=256$. In (d), $N=128$, 
    $W=4$ and $m=70$. In all cases $\chi=0.05.$}
    \label{fig2}
\end{figure}

In Fig. \ref{fig2}a we examine a typical scenario of structureless $Q_{\alpha\beta\gamma\delta}$ that displays a predicted RJ thermal 
distribution (dashed blue line). The main figure reports the final $|C_\alpha|^2$s (green circles) associated with a random initial 
preparation (black circles), demonstrating that, even though $Q_{\alpha\beta\gamma\delta}\sim 1/N^2$, a single nonlinear defect is able 
to thermalize the whole array. 

Next, we consider the case of structured $Q_{\alpha\beta\gamma\delta}$, which we achieve by inducing Anderson localization of the 
supermodes $f_\alpha$. This separates the modes into two distinct groups: those with high amplitude at the position of the nonlinear defect 
can interact with each other, while the rest are isolated. Due to the exponential localization, only modes that are centered in the proximity 
$\sim \ell_{\infty}$ of the nonlinear defect will belong to the first group. Combinations exclusively involving such modes will provide 
large $Q_{\alpha\beta\gamma\delta}$'s that will overpower the phase-matching mechanism. In practice, we have 
established a cut-off amplitude $|f^{\rm cutoff}(m)|^2\propto \langle\sum_l|f_\alpha(l)|^4\rangle_\alpha$ to identify these high-amplitude 
interacting modes. We have corroborated the selection of these modes by monitoring their modal power evolution.

This set of $N_p$ ``prethermal'' modes maintains (quasi-) constant internal energy ${\cal H}_P$ and power ${\cal A}_P$ for an exponentially 
long time (see Fig. \ref{fig2}b). This is suggestive of a thermodynamic analysis pertaining to a long-live metastable state. Applying 
the OT methodology to this subset we can extract from ${\cal H}_P, {\cal A}_P$ the corresponding $T_P,\mu_P$ defining a metastable 
RJ that dictates the power distribution among the prethermal modes. Importantly, as shown in Figs. \ref{fig2}c,d, 
the metastable RJ (solid blue line) might differ drastically from the final thermal RJ (dashed blue line). The open black circles 
and the solid green circles represent the initial and post-evolution power distributions respectively, evaluated via a direct numerical 
integration of Eq. (\ref{NLS_eq}). The prethermal modes that form the metastable RJ distribution are highlighted with red diamonds. 
We underline once more the longevity of these metastable states which renders them more practically relevant than the thermal RJ distribution.
This can be seen in Fig. \ref{fig2}b where the temporal evolution of the power occupations is shown for 
the longest time that we have computationally achieved. The formation of a metastable RJ is unique to cases of structured $Q_{\alpha
\beta\gamma\delta}$, as systems displaying structureless $Q_{\alpha\beta\gamma\delta}$ thermalize directly to the thermal RJ without 
undergoing an intermediate metastable configuration.

\begin{figure}[h!]
    \centering
    \includegraphics[width=0.45\textwidth]{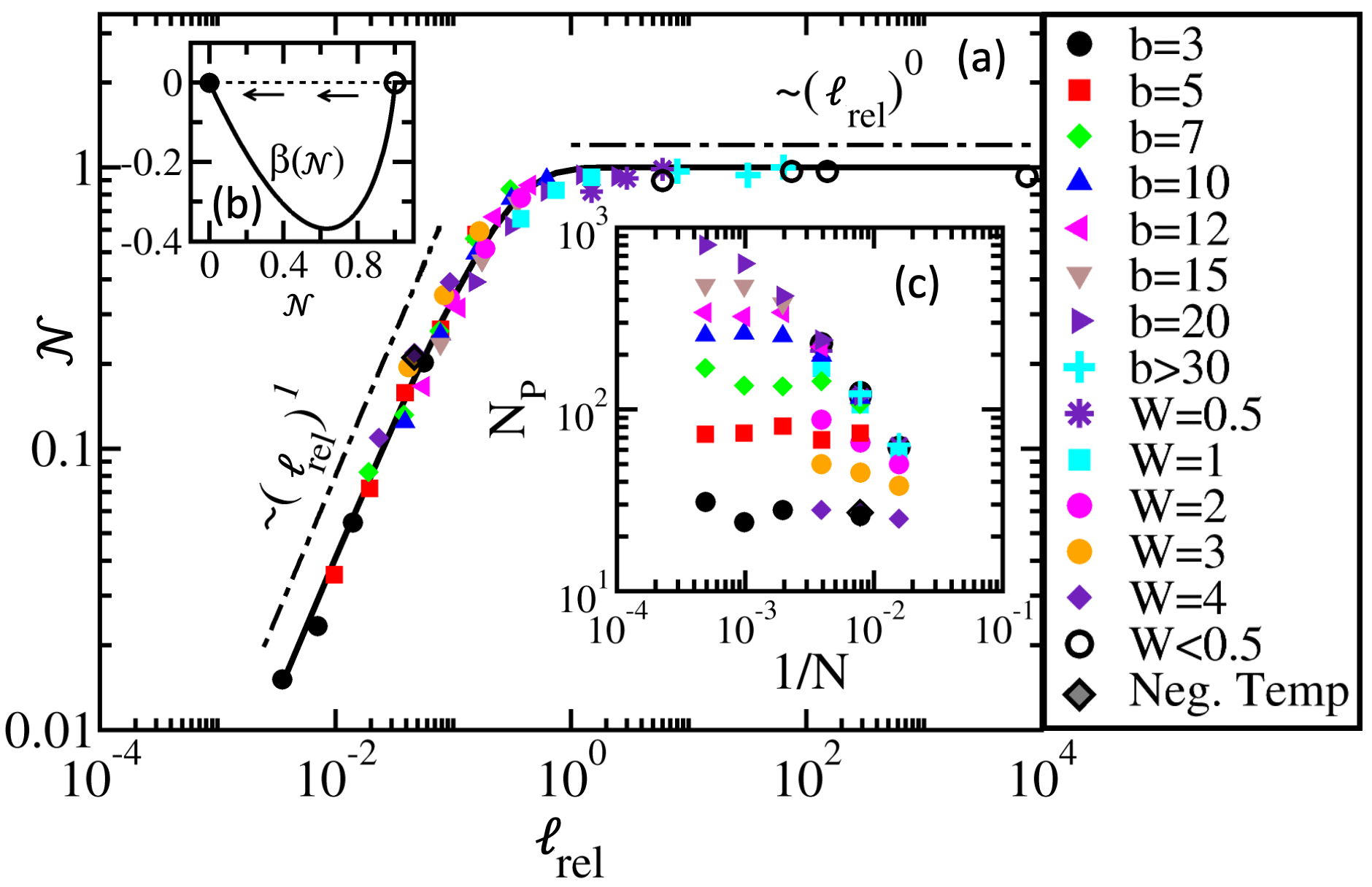}
    \caption{ (a) Prethermal mode density ($\cal{N}$), described by a one-parameter function 
    $\mathcal{N}(\ell_{rel})=1-e^{-5.4\ell_{rel}}$ that only depends on the relative localization length ($\ell_{rel}$). Data from 
    both NN and BRM systems are shown. Dot-dashed lines indicate the $\ell_{rel}^1$ and $\ell_{rel}^0$ behaviors. 
    (b) $\beta-$function associated with Eq. \ref{betafunction}, featuring two fixed points at $\mathcal{N}=1$ (unstable) and 
    $\mathcal{N}=0$ (stable). (c) Unscaled data showing the number of prethermal modes as a function of $N^{-1}$.}
    \label{fig3}
\end{figure}

{\it One parameter scaling and $\beta$-function formalism --} We are now equipped to formulate a one-parameter scaling theory that 
describes the number of prethermal modes $N_P$. The underlying ansatz is that, although the metastable RJ is determined by numerous 
system parameters (disorder configuration, connectivity, position of the nonlinear defect, $N$, ${\cal H}$, and ${\cal A}$), the 
number of prethermal modes is a simple function of the scaling variable $\ell_{\rm rel}\equiv l_\infty/N$. In other words, we 
postulate the existence of a universal function
\begin{equation}
    \mathcal{N}\equiv \frac{N_P}{N_{\rm ref}}
    = \Phi(\ell_{\rm rel})
        \approx \begin{cases}
        \ell_{\rm rel} & \text{for } \ell_{\rm rel}<0.1\\
        1&\text{for }\ell_{\rm rel}>1
    \end{cases},
    \label{master_curve}
\end{equation}
where $N_{\rm ref}\propto N$ is the number of prethermal modes associated with an underlying ``structureless'' (ergodic) system which 
acts as a reference system. For example, for BRM modeling, the reference ensemble is the full RMT $b=N$, where all 
modes are prethermal, i.e., $N_{\rm ref}=N$ and therefore ${\cal N}=N_P/N$ is the density of prethermal modes \cite{note}. 
We have numerically tested the validity of Eq. (\ref{master_curve}) by integrating Eq. (\ref{NLS_eq}) using both BRM and NN connectivity 
matrices $J_{ln}$. In the case of BRMs (NNs) various system sizes $64\leq N\leq 2048$ ($64 \leq N\leq 256$) and bandwidths 
$3\leq b \leq 64$ (disorder strengths $0.01\leq W\leq 4$) have been used. The numerical data shown in 
Fig. \ref{fig3} confirms the scaling ansatz Eq. (\ref{master_curve}), encapsulating the transition from completely thermalizing 
systems ($\mathcal{N}\rightarrow 1$) to systems that support metastable states ($\mathcal{N}<1$). Specifically, when the supermodes 
are extended over the whole system ($\ell_{rel}>1$), they all have a non-negligible amplitude at the nonlinear 
defect, and the entire system thermalizes. Conversely, when the localization length is smaller than the system size, ($\ell_{rel}<1$), 
the number of modes that interact with the nonlinear defect is $\sim \ell_\infty$, and thus, the prethermal modes will 
have a density $\mathcal{N}\sim \ell_{rel}$. A law valid in all regimes is $\Phi(\ell_{\rm rel})=1- 
e^{-C\cdot \ell_{\rm rel}}$ where $C=5.4$ is the best fitting parameter, see Fig. \ref{fig3}(a).

Following ideas from Renormalization Group Theory (RGT) we recast Eq. (\ref{master_curve}) into an equivalent form given by 
Eq. (\ref{betafunction}). This formulation highlights the fact that the density of prethermal modes is a solution of a 
(nonlinear) first-order differential equation. The resulting $\beta-$function takes the form $\beta(\mathcal{N})=(1-\mathcal{N})
\cdot \ln(1-\mathcal{N})$, which is always negative in the domain of definition $\mathcal{N}\in [0,1]$ (see Fig. \ref{fig3}(b)). 
Moreover, it is a continuous function since it describes how the density of the prethermal modes evolves as a function of
system size $N$.
Notice that, unlike the typical $\beta$-functions in RGT, our $\beta$-function is non-monotonic in $\mathcal{N}$ and features two 
fixed points, a stable one at $\mathcal{N}=0$ and an unstable one at $\mathcal{N}=1$. This is physically consistent with the fact 
that increasing the system size $N$ will decrease the fraction of prethermal modes $\mathcal{N}$.

{\it Relaxation Dynamics --} The dynamics towards the metastable and/or the thermal RJ state is described by the modal relaxation rates 
$\Gamma_\alpha$,
\begin{equation}
\label{gamma}
 \Gamma_{\alpha} = \frac{4\pi \chi^2}{n_\alpha} \sum^{\prime}_{\beta \gamma \delta} \left| Q_{\alpha \beta \gamma \delta}
 \right|^2 n_\beta \, n_\gamma \, n_\delta \,
 \delta(\varepsilon_\alpha +\varepsilon_\beta -\varepsilon_\gamma -\varepsilon_\delta),
\end{equation}
where $n_\alpha=T/(\varepsilon_\alpha-\mu)$. Equation (\ref{gamma}) was derived using a kinetic equation approach (see 
conditions for the derivation in Refs. \cite{SKS21,RSFCK23}). Its numerical evaluation utilized 
matrices of size $N=512, 1024$. For the statistical processing, we have used a sufficient number of disorder realizations 
to ensure that the total number of relaxation rates exceeded $35000$ data.

The form of $Q_{\alpha\beta\gamma\delta}$ allows us to reduce Eq. (\ref{gamma}) to 
\begin{equation}
\label{defectG}
\Gamma_{\alpha}\propto \chi^2 F_\alpha(T,\mu)\times \left|f_\alpha(m)\right|^2, 
\end{equation}
where $F_\alpha(T,\mu)$ incorporates the various mode-mixing terms appearing in the triple sum in Eq. (\ref{gamma}). When 
evaluating the relaxation process towards the metastable state, the summation involved in $F_\alpha(T,\mu)$ is restricted 
to the $N_P$ (quasi-) isolated prethermal modes.

\begin{figure}[h!]
    \centering
   \includegraphics[width=0.5\textwidth]{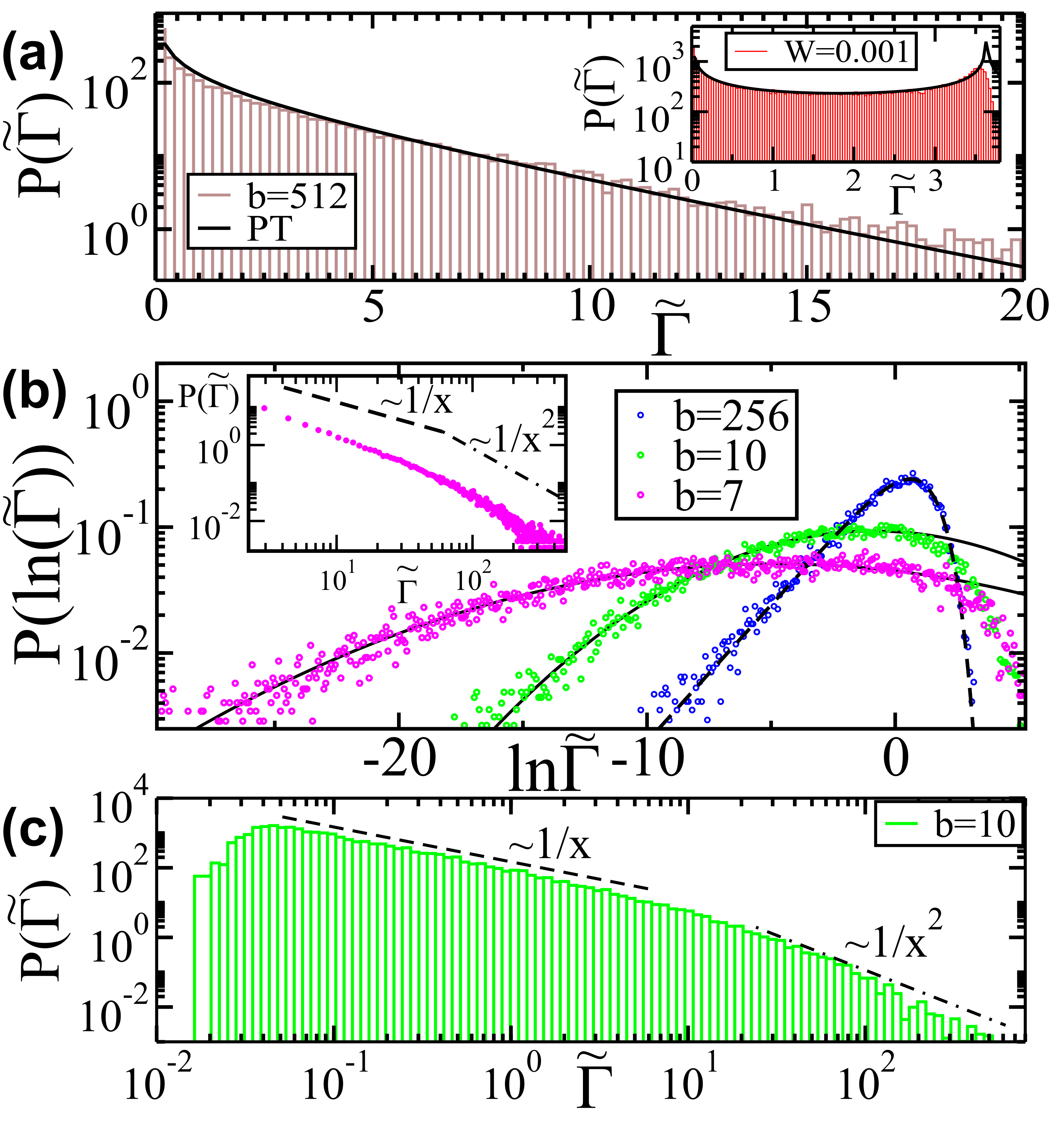}
    \caption{The distribution of relaxation rates ${\cal P}$ towards the thermal 
    RJ of (a) ballistic/chaotic and (b) disordered systems in the localized regime. Insets report (a) the distribution of a 1D 
    lattice with $W=0.001$ and (b) the intermediate regime of ${\cal P}(\tilde\Gamma)$ dominated by the prethermal modes. In (c) we report 
    ${\cal P}(\tilde\Gamma_P)$ of the (quasi-)isolated set of prethermal modes. The lines are the corresponding theoretical 
    predictions (see text). 
    }
    \label{fig4}
\end{figure}

When analyzing the distribution of relaxation rates ${\mathcal P}(\Gamma)$ towards the thermal RJ, we assume that the 
$\alpha-$supermode intensity at the position of the defect $\left|f_\alpha(m)\right|^2$ is the dominant statistical term 
in Eq. (\ref{defectG}). 
In linear wave chaotic structures, the 
distribution of supermode amplitudes is derived using Berry's random superposition hypothesis of plane waves \cite{E96,S99}.
Then, the probability density of intensities follows the Porter-Thomas (PT) distribution ${\cal P}(y=N|f_\alpha|^2)=(N/\sqrt{2\pi y}) 
e^{-y/2}$. From these considerations, we conclude that ${\cal P}({\tilde\Gamma}=N\Gamma)$ follows also a PT distribution. 
In Fig. \ref{fig4}a we report our numerical results together with the PT prediction. The observed agreement confirms 
our original assumption that the fluctuations of $F_\alpha$ do not manifest in the statistics of relaxation rates.

Peculiar behaviors might arise in cases where symmetries interfere with the underlying wave chaotic nature of a linear system. 
For example, in the limiting case $W\rightarrow 0$ of $1$D translational-invariant lattices, $f_\alpha(m)\sim N^{-1/2}\sin(k_\alpha m)$. 
Assuming uniformly distributed wavevectors $k_\alpha\in [-\pi,\pi]$, we get ${\mathcal P}(\tilde \Gamma \equiv \Gamma N)\sim 
[\tilde \Gamma (1-\tilde \Gamma/2)]^{-1/2}$. This is a bimodal distribution, 
indicating that there are two groups of fast- and slow- (or non-)thermalizing modes (see inset of Fig. \ref{fig4}a). 

When localization dominates the relaxation process, $\Gamma_\alpha\sim |f_\alpha(m)|^2\sim \exp(-2|x_\alpha-m|/l_\infty+
\eta_{x_\alpha})$. The stochastic variable $\eta_{x_\alpha}$ is Gaussian noise with zero mean and variance $\left(\Delta
\eta_{x_\alpha}\right)^2=(x_\alpha-m)/l_\infty$ that describes random fluctuations around the mean intensity profile 
\cite{IKPT97}. Consequently, ${\cal P}(\tilde\Gamma)\sim \exp\left(-(l_\infty/4N)\ln^2(\tilde\Gamma/\tilde\Gamma_0)\right)$ 
where $\ln(\tilde\Gamma_0)=-N/l_\infty$. This expression applies to the smallest relaxation rates associated with modes that are 
localized in a distance $|x_\alpha-m|\sim N$ from the defect. These predictions have been tested using both disorderd NN and 
BRM matrices (see Fig. \ref{fig4}b). 

Further analysis reveals an intermediate regime of $\Gamma-$values where the statistics is dominated by the presence of prethermal 
modes. Their centers of localization $x_\alpha$ are uniformly distributed in the proximity $\sim 
l_\infty$ of the defect site. Neglecting the $\eta_{x_\alpha}$ fluctuations, we get ${\cal P}(\tilde \Gamma)\sim 1/\tilde\Gamma$, 
see inset of Fig. \ref{fig4}b. For even larger relaxation rates, we employ a non-perturbative argument. It assumes that a number 
of modes in a distance $X=|x_\alpha-m|$ will relax at propagation distances $z\sim 1/\Gamma\sim X$. Since $X$ is proportional to 
the total measure $u\sim X\sim 1/\Gamma$ of these modes, it follows that ${\cal P}(\Gamma)=du/d\Gamma\sim 1/\Gamma^2$.

Similar arguments apply to the relaxation rates $\Gamma_P$ of the prethermal modes towards their metastable RJ. Repeating the same 
steps as above, we conclude that the distribution ${\cal P}(\tilde \Gamma_P)\sim 1/{\tilde \Gamma_P}$. The restriction of the summation 
to the $N_P$ prethermal modes gives a lower bound for $F_\alpha$ which in turn bounds $\Gamma_\alpha$. The numerical evaluation of 
the relaxation rates is shown in Fig. \ref{fig4}c and nicely confirms the above considerations.

Let us finally point out that the rescaled form of the relaxation rates ${\tilde \Gamma}$ that naturally appears in our 
analysis is reminiscent of the Thouless conductance defined as $g_T=\gamma/\Delta$ where $\Delta\sim 1/N$ is the 
mean level spacing and $\gamma$ is the linewidth of resonant modes. Like the Thouless conductance captures the disordered/chaotic 
nature of mesoscopic transport, our ${\tilde \Gamma}$ probes the underlying complexity of the linear systems in the thermalization 
process and reflects the transition from a ballistic to a localized behavior. 

\textit{Conclusions.-} We have analyzed the thermalization process of a beam propagating in complex nonlinear multimoded 
systems. We found that even a single nonlinear defect is capable of inducing thermalization described by an RJ thermal state. 
When Anderson localization effects are dominant, the mode-mode interactions acquire a nonuniform structure. This enforces 
the formation of metastable RJ states whose optical temperature and chemical potential can be dramatically different from 
the ones defining the thermal RJ state. The density of prethermal modes is determined by a universal $\beta-$function which 
describes the system without recourse to its microscopic details. These modes also have larger relaxation rates (towards the 
prethermal and/or final RJ) than the rest of the modes. We have analyzed the distribution of relaxation rates and identified 
the signatures of the transition from chaotic/
ballistic behaviors to localization. 


The use of a nonlinear defect as a ``probe" for Anderson localization (AL), is not at all trivial. Indeed, for the study of AL one needs to select states at a given energy E. However, in the thermalization problem the four-mode interaction mixes modes with different energies, so that even if the initial preparation selects a fixed energy, the subsequent evolution will create excitations in other energies. Further investigations along these lines are necessary in order to establish this criterion as a probe for AL.

\begin{acknowledgments}
We acknowledge support from the MPS Simons Collaboration via grant No. 733698. A.Y.R. and L.J.F.-A. acknowledge partial support from MINCyT and CONICET (Argentina) Grant No. CONVE-2023-10189190 - FFFLASH. We acknowledge useful discussions with Prof. B. Shapiro. 
\end{acknowledgments}




\begin{thebibliography}{30}
\bibitem{note}In cases of networks with symmetries, $N_{\rm ref}$ might be slightly smaller than the size of the system reflecting 
the underlying symmetries that prohibit the ergodicity of the supermodes. For example, NN networks with $W=0$ have supermodes 
whose nodal points may coincide with the position of the nonlinear defect, and thus, $Q_{\alpha\beta\gamma\delta}=0$. These modes 
would not participate in the thermalization process (even in the limit of infinite time). In such cases, one needs to assume 
the limiting value of $N_{\rm ref}$ as $W\rightarrow 0$. One can envision possibilities to exploit this scenario as protection 
against information melting.
\end{thebibliography}
\end{document}